\documentclass{elsart3p}

\usepackage[dvips]{color,graphicx}

\usepackage{amssymb}
\usepackage{units}

\newcommand{\ket}[1]{\left | \, #1 \right \rangle}

\begin{document}
\begin{frontmatter}
\title{Creating diamond color centers for quantum optical applications}

\author[oxford]{F. C. Waldermann\corauthref{cor}}
\corauth[cor]{Corresponding author; email: fe\-lix.\-wal\-der\-mann@phy\-sics.ox.ac.uk}
\author[melbourne]{P. Olivero}
\author[oxford]{J. Nunn}
\author[oxford]{K. Surmacz}
\author[oxford]{Z. Y. Wang}
\author[oxford]{D. Jaksch}
\author[oxford]{R. A. Taylor}
\author[oxford]{I. A. Walmsley}
\author[rmit]{M. Draganski}
\author[melbourne]{P. Reichart}
\author[melbourneCQCT]{A. D. Greentree}
\author[melbourneCQCT]{D. N. Jamieson}
\author[melbourneCQCT]{S. Prawer}

\address[oxford]{Clarendon Laboratory, University of Oxford, Parks Road, Oxford, OX1 3PU, UK}
\address[melbourne]{School of Physics,
The University of Melbourne, Parkville, Victoria 3010, Australia}
\address[rmit]{Applied Physics, RMIT University, GPO Box 2476V,
Melbourne, Victoria 3001, Australia}
\address[melbourneCQCT]{Center for Quantum Computer Technology, School of Physics,
The University of Melbourne, Parkville, Victoria 3010, Australia}

\date{09 March 2007}

\begin{abstract}
Nitrogen vacancy (NV) centers in diamond have distinct promise as
solid-state qubits.  This is because of their large dipole moment,
convenient level structure and very long room-temperature coherence times.
In general, a combination of ion irradiation and subsequent annealing is
used to create the centers, however for the rigorous demands of quantum
computing all processes need to be optimized, and decoherence due to the residual damage caused by the implantation process
itself must be mitigated.  To that end we have studied photoluminescence
(PL) from NV$^-$, NV$^0$ and GR1 centers formed by ion implantation of 2MeV He ions over a wide range of fluences. The sample was annealed at $600^{\circ}$C to minimize residual vacancy diffusion, allowing for the concurrent analysis of PL from NV centers and irradiation induced vacancies (GR1). We find non-monotic PL intensities with increasing ion fluence, monotonic increasing PL in NV$^0$/NV$^-$ and GR1/(NV$^0$ + NV$^1$) ratios, and increasing inhomogeneous broadening of the zero-phonon lines with increasing ion fluence.  All these results shed important light on the optimal formation conditions for NV qubits. We apply our findings to an off-resonant photonic quantum memory scheme using vibronic sidebands.
\end{abstract}


\end{frontmatter}

\section{Introduction}
The creation of single photoluminescence centers in diamond has
promising applications in the development of hybrid optical solid
state quantum devices. Diamond exhibits a large inventory of
optically active centers, related to the vibrational and electronic
states of impurities and defects in the crystal lattice
\cite{zaitsev_book}.  The prime candidate for many such applications
is the negatively charged nitrogen-vacancy (NV$^-$) center.  It has
proven to be a reliable source for single photons \cite{S_276_2012}
suitable for quantum key distribution \cite{PRL_89_1879011}, and
when used as a spin qubit has demonstrated coherent oscillations \cite{kcbld2003},
qubit tomography \cite{PRL_93_130501}, Stark shift control of the optical transition  and coherent population trapping \cite{sfsfbgodrrgrjp2006}.  Coupling of NV$^-$ centers
to nearby nuclear \cite{PRL_93_130501} and electron \cite{gdpwnjrsgpmthw2006,hmea2006}
spins has also been demonstrated. Furthermore, technology to sculpt
optically important nano and micro-structures in diamond is
progressing rapidly \cite{orrghrgsmjp2005}.  Based on such promising results, diamond seems a strong candidate platform for quantum information
processing, and several different schemes have been proposed,
including resonant dipole-dipole coupling,
repeat-until-success and brokered graph states \cite{hrb2004}.  A
recent review of this field can be found in Ref.
\cite{JPC_2006}.

The schemes mentioned above were all based on the formation and
control of \emph{isolated} qubits, however there are many protocols
which benefit from \emph{ensembles} of optically active centers. In
ensemble protocols, the emphasis is usually on using the system as a
medium to act on individual photons or photonic qubits, rather than
focusing on the centers themselves as the qubits.  Here we are
explicitly thinking about protocols such as the weak nonlinear
coupling schemes \cite{mns2005}, or photon storage schemes \cite{dlcz2001,nwrswwj2007}.  Again, NV$^-$ appears to be an extremely suitable scheme for
such applications, and electromagnetically induced transparency has been demonstrated at radio-frequency \cite{wm1999} and optical
transitions \cite{sfsfbmsgodrrgrhjp2006} in ensembles.  Our principal concern in the present work is the optimization of the formation methods of ensembles for these applications, with special focus on storage of
light outlined in Ref.~\cite{nwrswwj2007}.

Beyond the obvious attraction of ensembles in increasing the density
of particles over the single particle case, ensembles of emitters
have other advantages.  The electromagnetic interaction strength is
strongly enhanced \cite{lfcdjcz2001}, which increases the fidelity
of state transfer and storage of quantum information
\cite{snwwwj2006}. Of great pertinence to quantum memory schemes is
that the distributed storage of quantum information provides more
temporal stability. Moreover, ensembles account for the modal
structure of interacting light and allow for the storage of the
temporal shape of qubit photons \cite{nwrswwj2007}.

The realization of scalable devices for the above-mentioned
applications requires adequate control of the formation processes of
active NV centers in diamond crystals.  As the final applications of
the NV centers differ from the previous motivations (which were
usually into the fundamental properties of the centers), so too are
the formation requirements.  In particular, for quantum device
applications, we require the ability to engineer spatially defined
ensembles, perhaps in some patterned array, with minimal residual
inhomogeneity, and compatibility with sculpted optical structures.
These are demanding requirements indeed, and force a critical
reappraisal of the existing methodologies.

At least two alternative routes are available for the creation of NV
centers in single crystal diamond (a third, where NV centers are
created during diamond growth is more pertinent to diamond formed by
chemical vapor deposition, e.g. \cite{APL_88_023113}).  Firstly radiation
damage from various sources (electrons, neutrons, ions, etc.) can be
used to create vacancies in nitrogen-rich (i.e. type Ib) diamond, which combine to form NV post-anneal.  Secondly, direct
implantation of nitrogen in the purest (i.e. type IIa) crystals, where the NV center is formed by the vacancies due to the N
implantation. While the latter strategy represents a suitable method
to fabricate low density NV ensembles or isolated qubits in diamond,
the first strategy provides a ready method to quickly and
efficiently produce high density ensembles.  Because of our focus on
ensemble protocols, this is the approach that we have chosen in this
study.  Here, we present an extensive study on the effects of MeV
He$^{+}$ ion implantation in type Ib diamond crystals, as a
candidate technique to create high-density NV ensembles in diamond.
Irradiation with high energy ions creates vacancies as the ion
looses energy to the lattice.  Such processes can be easily and
accurately modeled using the standard TRIM package \cite{trim}.  Because the energy loss of the ions is
non-uniform with distance into the material, with most damage
created at the end of range of the ion, this perforce leads to a
highly non-uniform depth profile of NV centers formed post-anneal.

Although NV ensembles in diamond are promising for optical quantum information processing (QIP),
there are several material-related issues that need to be fully
understood if ion irradiation is to be considered for the creation
of NV center ensembles:

\textbf{NV center density.} A high conversion efficiency from
nitrogen and vacancy defects to optically active NV centers is
crucial to reach high NV densities, increasing the probability of an
interaction, but also minimizing the number of unconverted N which
will contribute to decoherence \emph{without} contributing to the
interaction. An active NV center can be created in a diamond crystal
containing nitrogen and vacancy defects with thermal annealing at
temperature $\geq600^{\circ}$C. At such temperatures
\cite{DRM_7_228}, the vacancies start migrating to the nearest
substitutional nitrogen atoms, where their aggregation is
energetically favorable \cite{JPCM_13_6015}. The conversion
efficiency is limited due to competitive processes such as the
formation of other defects and vacancy-interstitial recombination.

\textbf{Temporal stability of the NV center.} The NV defect can
exist in two charge-states (NV$^{-}$ and NV$^{0}$) that have been correlated with $\lambda$=638 nm and $\lambda$=575 nm luminescence
emissions, respectively  \cite{zaitsev_book}. Charge transfer
\cite{DRM_14_1705,JPCM_12_189,JPCM_12_7843,axcm0508323} mechanisms
lead to the conversion between NV$^{-}$ and NV$^{0}$ centers
(photo-ionization). This process  appears experimentally as
photo-bleaching. The equilibrium between NV$^{-}$
and NV$^{0}$ concentrations is determined by the presence of
nitrogen donors in the lattice  \cite{wscb2003,JPCM_14_3743}. Only
the NV$^-$ center has an easily accessible $\Lambda$-shaped energy
desired for many quantum computation applications, and a high NV$^-$
/ NV$^0$ ratio is desirable.

\textbf{Inhomogeneous spectral broadening.} The inhomogeneous
broadening of the zero phonon line (ZPL) NV$^{-}$ emission is
usually reported as 750 GHz \cite{zaitsev_book}, which significantly
exceeds the homogeneous linewidth corresponding to a transition
lifetime $\sim$12~ns. Such large broadening is commonly attributed
to variations in strain and electric fields within the diamond
crystal, due both to impurities and structural defects
\cite{JPC_2006}, although recent work in low N diamond where NV
concentrations less than the canonical 750~GHz have been observed
seem to indicate that this linewidth can be altered by suitable
preparation. Inhomogeneous broadening can lead to dephasing or
undesired transitions in resonant QIP schemes, while off-resonant
\cite{nwrswwj2007} or frequency-selective \cite{sfsfbgodrrgrjp2006}
schemes are less vulnerable to it.

\begin{figure}[tbp]
    \centering
        \includegraphics[height=6.2cm]{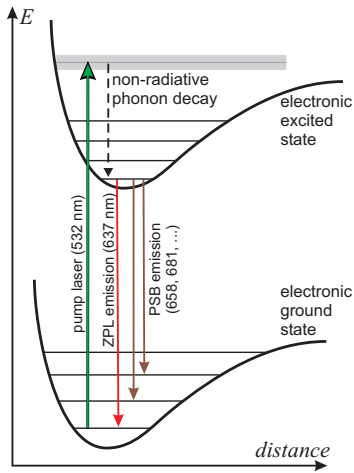}
    \caption{Typical emission scheme for a color center in diamond for short wavelength excitation and at cryogenic temperatures. After excitation, fast non-radiative decay transfers the population to the vibrational ground state of the electronic excited state. From here, radiative decay occurs both into the zero phonon line (ZPL) and the phonon sidebands (PSBs).}
    \label{fig:NVsidebandemission}
\end{figure}

\begin{figure}[tbp]
    \centering
        \includegraphics[height=6.2cm]{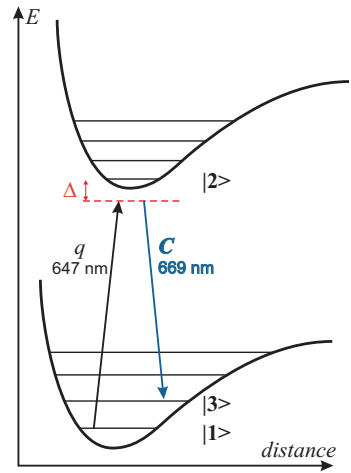}
    \caption{Off-resonant storage scheme of a broadband photon $q$ into a local phonon mode of a defect center, for example, the NV$^-$ center. $\Delta$ denotes the detuning of the transition from resonance, $C$ is the classical control pulse. The actual storage is achieved in an ensemble of absorbers. \quad \quad \quad \quad \quad \quad \quad \quad \quad \quad \quad \quad \quad \quad \
     }
    \label{fig:NVsideband}
\end{figure}

The above mentioned issues are related with the fabrication of NV
centers in diamond. They might impose limits on the suitability of
NV centers for certain QIP schemes. In the present work, we compare
our findings on NV center ensembles with the requirements for a
photonic quantum memory scheme (Qmem) that uses broadband (i.e. sub-picosecond) pulses and an off-resonant transition  \cite{nwrswwj2007}, often called Raman quantum memory \cite{gafsl2007}. This scheme utilizes
a strong classical control pulse to absorb the qubit photon into a
storage state in the material, which is a collective state of the
ensemble. Unlike the typical QIP schemes that use the spin-state of the NV center, here, the first excited phonon state has been chosen as
the storage state (see figure \ref{fig:NVsideband}). Most optically active defect centers in diamond show phonon sidebands (PSBs) \cite{zaitsev2000}, and PSB storage as introduced here could in principle be realised in many defect centers. The thermal population of this state is $10^{-80}$ at $T\sim 4$ K, making state preparation, e.g. by optical pumping, unnecessary. The read-out of
the stored photon can be achieved using another strong classical
control pulse. 

Raman-like transitions are typically insensitive to inhomogeneous
broadening of the upper level, provided that the energy separation
between ground and storage state is well-defined throughout the ensemble.
Certainly this is the case for the phonon sidebands considered here.
However, with increasing inhomogeneity, a larger detuning is necessary and a stronger control field is necessary to reach a sufficient strength of the two-photon transition. On the other hand, a large detuning allows for the off-resonant use of truly broadband pulses. For the NV$^-$ center, the storage state splitting is $\omega_{13} = \unit[15.3]{THz}$, allowing for a bandwidth of $\delta \sim \unit[3]{THz}$, i.e. sub-picosecond photon storage. Due to the high refractive index of diamond, the storage is optimum for irradiation at the Brewster angle, to avoid reflection-loss of the qubit photon.

Our scheme is not affected by the strong phonon sideband emission typical for the NV center, as the transition is truly off-resonant and the excited state is never populated. This is obvious at storage, as the transition from state $\ket{2}$ to $\ket{3}$ occurs stimulated by the strong control field and thus only into the one desired PSB. At readout, emission of a \emph{single} NV center would occur into all phonon sidebands, which would lead to a non-deterministic frequency of the recreated photon. While this affects Qmems based on single NV centers (and additional measures have to be taken such as suppression of PSB emission by microcavities), this problem is automatically avoided by an \emph{ensemble} quantum memory. Due to collective enhancement of ensemble emission, the qubit photon is emitted at the ZPL frequency only. For any other transitions (PSB transitions), no phase-matching \cite{nwrswwj2006arxiv} can be achieved, effectively suppressing the emission into these modes. 

Taking into account the practical considerations for light storage
in a realistic sample of NV diamond, we find that at a fluence
$\mathrm{F}=1\times10^{15}$ ions cm$^{-2}$, a $\lambda = 648$ nm
photon with picosecond bandwidth could be stored using a 16 $\mu$W
ultrafast control laser at 80 MHz repetition rate (transition occurs
at $\Delta \lambda = \unit[10]{nm}$ detuning from the ZPL resonance). With these parameters, the storage and retrieval fidelities would range in the percent region, and a proof of principle experiment could be established. Higher Qmem fidelities require higher optical depths, which could be achieved by different implantation techniques, e.g. multiple energy ion implantation or high energy electron implantation. It has to be stressed that, although we calculate this scheme for the
NV$^-$ center only, it can in principle be applied to any optically
active defect in diamond with a $\Lambda$-shaped energy level
structure, for example the GR1 defect, which also has an optically
accessible phonon-sideband.  This implies that these phononic
schemes for light storage are considerably more flexible than the
more conventional approach using existing atomic or molecular
levels. The current sample therefore has been annealed at a comparatively low temperature (see next chapter), to allow for the PL-analysis of both NV and GR1.

\section{Experimental Setup}
\subsection{Implantation and Annealing}
The employed sample was an artificial diamond produced by Sumitomo. The crystal was cut and polished from a large single crystal which was synthesized under high pressure and temperature (HPHT). All sides of the sample have [100] face orientation, and the crystal is classified as type Ib; the nitrogen concentration is $\sim100$ ppm, as reported by the manufacturers. All irradiated regions were in the central [100] growth sector of the sample, where negligible fluctuations in the nitrogen concentration can be expected, as verified by cathodoluminescence mapping. 

\begin{figure}[htbp]
	\centering
	\includegraphics[width=1.0\columnwidth]{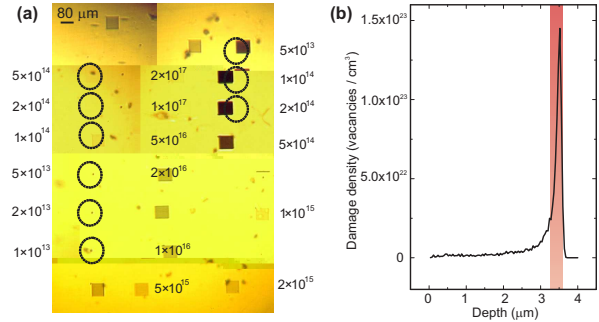}
	\caption{
	\textbf{(a)} Transmission optical microscope image of the implanted sample; the 80$\times$80 $\mu$m$^2$ implanted regions are labeled with the relevant ion fluences (ions cm$^{-2}$). Regions implanted at higher fluences are clearly visible because the damaged layer is opaque, while regions implanted at fluences below $10^{15}$ ions cm$^{-2}$ are not visible optically.
	\textbf{(b)} TRIM Monte Carlo simulation of the damage density profile induced in diamond by 2~MeV He$^+$ ions, scaled for an implantation density of $1 \times 10^{17}$ ions cm$^{-2}$; the damage density is concentrated at the end of the range, namely {${\sim 3.5 \ \mu\mathrm{m}}$} below the sample surface.
	}
	\label{fig:samplemicroscope}
\end{figure}

The sample was implanted with 2 MeV He$^{+}$ ions on the MP2 microbeam line of the 5U NEC Pelletron accelerator at The University of Melbourne. Regions of $80\times80 \ \mu\mathrm{m}^{2}$ were implanted in the central 100 growth sector of the sample at fluences ranging from $1\times10^{13}$ ions cm$^{-2}$ to 2$\times$10$^{17}$ ions cm$^{-2}$, using a raster scanning ion beam which was focused to a micrometer-sized spot, providing homogeneous ion fluence. The ion beam current ranged from $\sim$5 pA to $\sim$0.5 nA for different implantations, with implantation times $\lesssim 10 - 30$ min. The ion fluence was measured by monitoring the Rutherford back-scattered ions, after having coated the sample with carbon in order to avoid charging; afterwards, the conductive coating was removed by chemically etching. The sample was tilted by an arbitrary angle in order to avoid channeling effects.

Figure \ref{fig:samplemicroscope}a shows a transmission microscope
image of the sample after ion implantation; the regions implanted at
higher fluences are clearly visible because the high damage density
makes the material opaque, while at fluences below
1$\times$10$^{15}$ ions cm$^{-2}$, the implanted areas are invisible
to optical microscopy. Figure \ref{fig:samplemicroscope}b shows the
vacancy density profile of 2 MeV He$^{+}$ ions in diamond,
calculated with ``Transport of ions in matter'' (TRIM) Monte Carlo
simulation code \cite{trim}. The simulation takes into account both
direct collisions and secondary recoil events, and the value of the
atom displacement energy was set to 50 eV. Ion-induced damage causes
a progressive amorphization of the crystalline structure with
increasing ion fluences, which is localized mainly at the end of
range of implanted ions (namely $\sim 3.5 \ \mu\mathrm{m}$ below the
sample surface, as shown in Fig. \ref{fig:samplemicroscope}b), where
most of the nuclear collisions occur.

The evolution of the amorphized structure upon thermal annealing
depends critically on the damage density. In regions where the
vacancy density is above a critical threshold, the diamond
crystalline structure is permanently converted to a sp$^{2}$-bonded
phase. When the damage density is below the critical threshold,
thermal annealing has the effect of converting the amorphized
structure back to the crystalline diamond phase, although residual
point defects can form in the crystal. During the thermal annealing,
vacancies (interstitials) start becoming mobile in the crystal
lattice at temperatures of $\sim 600^{\circ}$C \cite{DRM_7_228}; in
this process, they can either recombine with interstitial atoms, or
form a vast range of luminescent centers, both isolated and paired
with native impurities (N, B, Si, Ni, etc.)  \cite{zaitsev_book}.

The sample was annealed for 1 hour in forming gas (4\% hydrogen in argon) atmosphere. The sample was annealed at the relatively low temperature of $600^{\circ}$C, which represents the onset for vacancy migration in diamond. This low annealing temperature induces only a small mobility of the isolated vacancies, avoiding a strong reduction in their concentration, and only a small fraction of them migrate to substitutional nitrogen atoms where they combine to form NV centers. This allows for the concurrent presence of both NV$^-$, NV$^0$ and GR1 centers in the material, whose photoluminescence analysis is the aim of this work. 
Using these TRIM calculations, the He ion fluence has been converted into the average vacancy density of the 3-$\mu$m-thick cap layer before annealing. After the annealing less than 0.25\% of the vacancies is expected to anneal out \cite{PRB_46_13157}, thus making our estimation of the damage-induced vacancies reasonably accurate also after the annealing processing. The cap layer is defined as the first 3 $\mu$m of the sample, as shown in figure \ref{fig:samplemicroscope}b. While this calculation is a good estimate at low ion fluences, it has to be noted that it is approximative only at high fluences due to non-linear material processes scaling (e.g. graphitization).

\subsection{Optical Experiments}
PL experiments were carried out using a Renishaw Raman apparatus with a Leica DMLM microscope. The sample was cooled to $T\sim4$ K in an Oxford Instruments liquid helium flow cryostat, which could be positioned with micrometer precision. A reflecting microscope objective was employed for both excitation and signal collection, minimizing the effects of chromatic aberration. The lateral resolution of the system was $\sim 2 \ \mu$m and therefore much smaller than the size of the implanted regions. The photoluminescence was induced by $\lambda_{\mathrm{exc}} = 532$ nm light from a Suwtech single longitude mode laser.

\begin{figure}[htbp]
	\includegraphics[width=1.0\columnwidth]{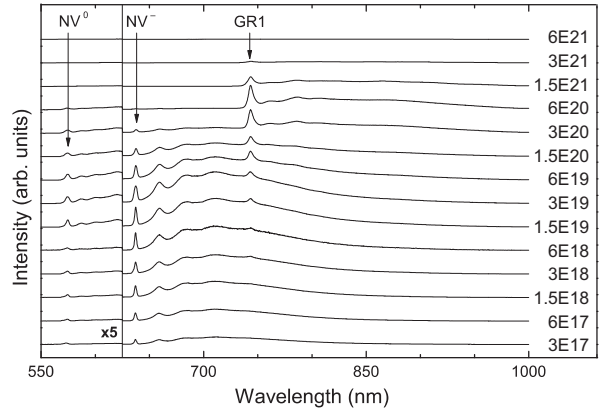}
	\caption{PL spectra for various irradiation fluences. The ZPLs of the color centers have been indicated by arrows, their PSB are to the right. The left end of the spectra, containing the NV$^0$ peak, is magnified by a factor of 5. The vacancy density (cm$^{-3}$) due to ion implantation in the cap layer (see text) is denoted on the right. }
	\label{fig:plregionsext}
\end{figure}

All spectra were acquired with the excitation and collection focus on the sample surface. Since the depth profile of the induced damage is non-uniform, slightly different focusing conditions might in principle affect the intensity of the PL signal; normalization to the intensity of the first order diamond Raman line is not possible since this is also significantly affected by ion-induced damage. Therefore, the reproducibility of the focusing conditions were tested by repeated measurements at various sample positions, from which variations of only $2\%$ were observed. Our spectra were recorded using the spectrometer with 300 and 1200 grooves/mm diffraction gratings. With $\lambda_{\mathrm{exc}} = 532$ nm laser excitation, the NV$^0$ PL emission and first order Raman line (Raman shift=1332 cm$^{-1}$, corresponding to $\lambda_{\mathrm{Raman}} = 572$ nm) partially overlap. The Raman emission has been subtracted from the NV$^0$ PL signal. While an absolute estimate of the color center densities from the photoluminescence signal is difficult and errorprone, we can derive relative trends of the center densities using the relative intensities of PL signals acquired at the various implanted regions of the sample, which is necessary for optimizing the conditions for NV formation.

\section{Results}
Figure \ref{fig:plregionsext} shows PL spectra from regions implanted at 14 different ion fluences, ranging from $2\times10^{17}$ ions cm$^{-2}$ to $1\times10^{13}$ ions cm$^{-2}$. The NV$^-$ and GR1 peaks ($\lambda$=638 nm and $\lambda$=742 nm, respectively) are clearly visible for most regions, although the GR1 doublet structure is not resolved due to an overlap of the two peaks attributed to inhomogeneous broadening. The NV$^{0}$ emission is at $\lambda$=575 nm and much weaker, but visible for some regions. The broad peaks on the red side of each ZPL peak arise from vibronic emission (phonon side bands).

\begin{figure}[tbp]
	\includegraphics[width=1\columnwidth]{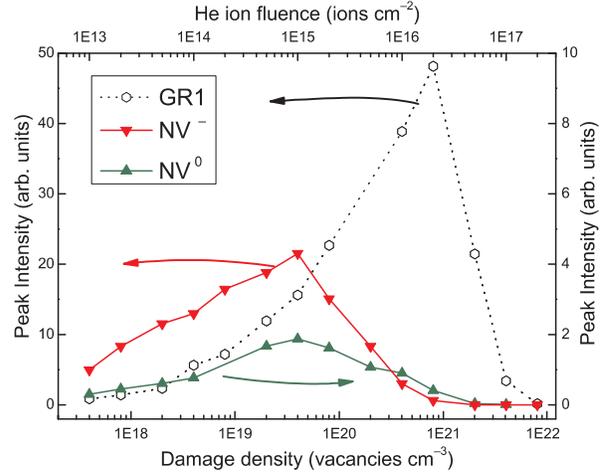}
	\caption{
	Intensity of the NV$^{-}$, NV$^{0}$ and GR1 zero-phonon-line PL as a function of the implanted ion fluence (upper X scale) and corresponding vacancy density in the cap layer (lower X scale; see text for definition). The NV and GR1 emission intensities have a non-monotonic behavior, exhibiting maxima at $3\times10^{19}$ vacancies cm$^{-3}$ and $6\times10^{20}$ vacancies cm$^{-3}$ for NV and GR1, respectively.} 
	\label{fig:plintensity}
\end{figure}

\subsection{PL intensities as a function of ion fluence}
Fig. \ref{fig:plintensity} shows the intensities of the ZPL emissions from NV$^{-}$, NV$^{0}$ and GR1 centers as a function of the implanted ion fluence. In all cases, the PL intensities follow a non-monotonic trend. After a steady sub-linear increase, the NV$^{-}$ and NV$^{0}$ emissions reach a maximum at fluence $\mathrm{F}=1 \times 10^{15}$ ions cm$^{-2}$, then steadily decrease until maximum fluence $\mathrm{F}=2\times10^{17}$ ions cm$^{-2}$ is reached. In interpreting this trend, it is necessary to consider the number of vacancies created by implanted He$^{+}$ ions at various fluences, as compared with the native substitutional nitrogen concentration of $\sim2\times10^{19}$ cm$^{-3}$, corresponding to a 100 ppm concentration. By integration of the vacancy profile reported in Fig. \ref{fig:samplemicroscope}b, one can estimate that each 2 MeV He$^{+}$ ion on average creates 38 vacancies in the diamond crystal. The vacancy density profile is strongly non-uniform with depth, with $\sim$~80$\%$ of the vacancies being created in the last 0.5 $\mu$m of the total 3.5 $\mu$m ion damage range (highlighted in Fig. \ref{fig:samplemicroscope}b). 
As visible in the transmission microscope image in Fig. \ref{fig:samplemicroscope}a, regions implanted at fluences $\mathrm{F}>1\times10^{15}$ ions cm$^{-2}$ turn opaque because of graphitization in the buried heavily-damaged layer $3.5 \ \mu$m below the diamond surface. Therefore, at high ion fluences only the $\sim3 \ \mu$m-thick surface layer is responsible for the enhanced PL signal, the end-of-range layer being optically opaque. 

At fluence $\mathrm{F}=1\times10^{15}$ ions cm$^{-2}$, about $3\times10^{19}$ cm$^{-3}$ vacancies are created in the surface layer with a relatively uniform depth distribution. As this vacancy density exceeds the nitrogen density ($\sim2\times10^{19}$ cm$^{-3}$), the increase of NV PL with increasing irradiation fluence stops at $\mathrm{F} \gtrsim 1\times10^{15}$ ions cm$^{-2}$. However, due to the low annealing temperature, not all vacancies associate with nitrogen atoms (see below). 

\begin{figure}[tbp]
	\includegraphics[width=0.8\columnwidth]{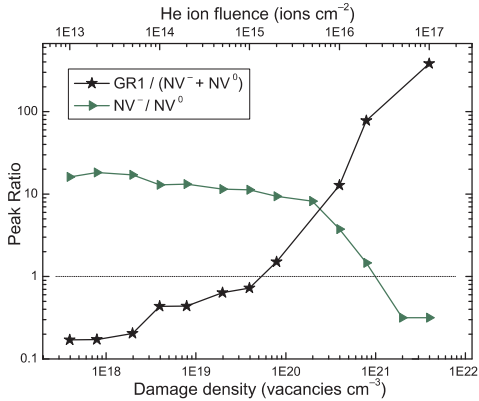}
	\caption{
	Ratios of photoluminescence intensities. At about $3 \times 10^{20}$ vacancies cm$^{-3}$, the fraction of NV$^-$ decreases due to a lack of free negative charge in the material, provided by substitutional nitrogen atoms (see text).}
	\label{fig:plratio}
\end{figure}

A confirmation for this interpretation can be found in the increase of the relative intensity of the GR1 emission with respect to the NV emission, as shown in Fig. \ref{fig:plratio}. The GR1 emission is unambiguously attributed to isolated neutral vacancies in the diamond structure  \cite{zaitsev_book}, and such a trend clearly indicates that isolated vacancies progressively exceed the NV centers population. It is worth remarking that GR1 emission is measured at low (i.e. $<1\times10^{15}$ cm$^{-2}$) ion fluences; this indicates that even where the native substitutional nitrogen concentration is higher than the vacancy concentration, not all the created vacancies get trapped at an N atom during thermal annealing. Again, this is due to a limited efficiency in the N+V$\rightarrow$NV conversion process \cite{APL_88_023113}, as is expectable for the relatively low annealing temperature used.

As shown in Fig. \ref{fig:plintensity}, the NV$^{-}$ and NV$^{0}$ intensities, after having reached a maximum, start decreasing at higher ion fluences. While at first a plateau behavior might be expected after the saturation of native nitrogen at $\mathrm{F}\sim 1\times10^{15}$ ions cm$^{-2}$, it must be considered that the increasing damage density is detrimental to the NV emission, both in terms of the increasing optical absorption of the damaged crystal and of the competitive formation of other luminescent centers. Damage induced absorption is qualitatively visible in the optical microscope image (figure \ref{fig:samplemicroscope}a) from regions implanted at fluences higher than $\mathrm{F}=1\times10^{15}$ ions cm$^{-2}$.

At high ion fluences, the strong suppression of NV PL detected indicates that graphitization starts affecting even the cap layer. Although it is likely that the NV density increases in the material, most of the NV centers are not optically accessible any more. 

For the annealing process employed here, the optimum damage density for the formation of optically accessible NV ensembles in type Ib diamond by He ion implantation is $\sim3\times10^{19}$ vacancies cm$^{-3}$; such a density can be achieved with medium ion implantation fluences ($\sim1\times10^{15}$ cm$^{-2}$ for 2 MeV He$^{+}$ ions). As a higher annealing temperature would increase the vacancy mobility and therefore revert the material damage to some extend, a higher annealing temperature is expected to shift the NV peaks in figure \ref{fig:plintensity} to the right. Annealing temperatures of $800^{\circ}$C have been reported to maximize the density of optically active NV centers \cite{JPCM_12_7843}, while the GR1 PL would be strongly reduced at higher annealing temperatures \cite{PRB_46_13157}. The GR1 emission also shows a non-monotonic trend, with a maximum at $\mathrm{F}=2\times10^{16}$ ions cm$^{-2}$. Again, the abrupt decrease of the GR1 intensity at higher fluences can be explained with the damage-induced optical absorption.

Assuming a $\sim 0.25\%$ efficiency in the conversion from nitrogen-vacancy pairs to optically active NV centers upon thermal annealing, we can estimate a density of optically active NV centers of $10^{17}$ cm$^{-3}$ in the cap layer. This corresponds to an optical depth of $D = 0.01$, while for higher annealing temperatures $D \sim 0.2$ could be reached due to a higher conversion efficiency, corresponding to a number of centers of $N_{\mathrm{NV}^-} = 5 \times 10^6$ in a $\unit[1]{\mu m}$ confocal spot. Here, the optical depth has been calculated from 
\[
D = \frac{1}{2 \hbar \epsilon_0 c} 
\ \frac{d^2 \nu_{12} \sigma}{\gamma},
\] 
where $\nu_{12}$ denotes the ZPL frequency, $\gamma$ the decay rate of the excited state, $\sigma$ the surface density of absorbers, and $d$ the dipole moment (estimated from the radiated lifetime $T_1 = \unit[12]{ns}$ as $d = [3\pi \epsilon_0 \hbar c^3 \gamma f_{12} / \omega^3]^{1/2} \sim \unit[5.5\times 10^{-30}]{Cm}$, where $f_{12}$ is the fraction of state $\ket{2}$ decay that goes into the ZPL). 

\begin{table}[htbp]
	\centering
	\begin{center}
\begin{tabular}{|r @{} l|r @{} l|c|}
\hline
 $\eta$ & (\%)\	& \ {$C$} & & $Q$ \\ \hline \hline 
 \ 17& 			& \ 0&.2 & \ \ 1 (no cavity)\ \ \\ \hline 
 \ 90& 			& \ 9&.0 & 79 \\ \hline 
 \ 99& 			& \ $10$&$^2$ & $8.7 \times 10^2$ \\ \hline 
 \ 99&.9  		& \ $10$&$^3$ & $8.7 \times 10^3$ \\ \hline 
 \ 99&.99 		& \ $10$&$^4$ & $8.7 \times 10^4$ \\ \hline 
\end{tabular}
\end{center}

	\caption{Qmem fidelity $\eta$ of storage or retrieval for an NV$^-$ ensemble Qmem, with necessary cooperativity parameters $C$ and microcavity $Q$-factors. An implantation density of $F=10^{15}$ ions cm$^{-2}$ and a N $\rightarrow$ NV conversion efficiency of $5\%$ for the cap layer have been assumed. }	
	\label{tab:qmemfidelities}
\end{table}

In the context of the off-resonant Qmem scheme \cite{nwrswwj2007} introduced above, a high ensemble density is necessary to reach a high fidelity. A generic result derived by Gorshkov \cite{gals2007i} covering all three types of Qmems (Raman, EIT and CRIB) states that the fidelities for storage and retrieval each are given by 
\[
\eta = \frac{C}{C+1}, 
\]
where $C \sim D$ is the cooperativity parameter, and $D$ the optical depth. Thus, for NV densities corresponding to a fluence of $\mathrm{F}=1\times10^{15}$ ions cm$^{-2}$, and optimised annealing temperature of 800 degrees, a Qmem would have an efficiency of $\eta \sim 17\%$, which is sufficient for a proof of concept. Practically, sufficient control field energy would be easy to achieve. Assuming an 80 MHz  repetition rate and a micro-PL setup, the transition could be realized for a control-field laser average power of 1~mW at a $\Delta\lambda = \unit[10]{nm}$ detuning from the ZPL resonance. However, for realistic QIP applications, a much higher storage fidelity is necessary, which for ion implantated samples could be reached by a longer propagation length, e.g. through a waveguide-structure \cite{orrghrgsmjp2005}, or by an enhanced light-matter interaction due to a microcavity. Here, a microcavity of much lower $Q$-factor would suffice, e.g. $Q \sim 1000$, compared to $Q$-factors several orders of magnitude higher that are necessary for high fidelity Qmems based on single NV centers. 

\subsection{Ratio of NV$^-$ and NV$^0$ PL}
Few studies have been conducted on the relative  $\mathrm{NV}^-/\mathrm{NV}^0$ intensity as a function of induced damage, using radiation sources (neutrons, electrons) that do not allow a straightforward estimation of the induced damage. In the context of QIP with NV$^-$ centers, a very high ratio is desirable, as the occurrence of NV$^0$ disturbs optical transitions involving NV$^-$ centers due to a spatial variation of the refractive index and temporal instability. This will be explained in more detail in the next section. 

A common characteristic of the spectra reported here is the low intensity of the NV$^{0}$ emission with respect to the NV$^{-}$ and GR1 emission. This is due to the fact that the type Ib diamond sample has a relatively high concentration of single substitutional nitrogen. Nitrogen acts as an electron donor in diamond with an ionization energy of 1.7 eV, while the ground state of of an NV$^{-}$ center lies 2.58 eV below the conduction band \cite{scdg2000}. 
The electrons provided by nitrogen donors are responsible for the conversion of nearby NV$^{0}$ centers to the charged NV$^{-}$ state  \cite{JPCM_14_3743}; the process can be described by the reaction: ${\mathrm{NV}}^0 + {\mathrm{N}} \rightleftharpoons {\mathrm{NV}}^- + {{\mathrm{N}}}^+$. Studies carried out on samples with different concentrations of substitutional nitrogen proved that at low concentration, the NV emission arises mainly from the neutral centers, while at high concentration, NV$^-$ luminescence becomes largely predominant \cite{wscb2003,kcbld2003}. This is explained by statistical arguments with a variation of the Fermi level as a function of N doping concentration  \cite{PRB_53_11360,JPCM_12_189,JPCM_12_7843}, or microscopically with the proximity of the NV defect to a nitrogen donor  \cite{JPCM_14_3743,axcm0508323}. Our $\mathrm{NV}^-/\mathrm{NV}^0$ ratio of about 10 for low and medium implantation densities is consistent with previous reports \cite{wscb2003}, measuring an $\mathrm{NV}^-/\mathrm{NV}^0$ ratio of 3.3 for a CVD grown sample with a medium-high nitrogen density of 47 ppm. As the photoluminescence excitation efficiency most likely varies for NV$^0$ and NV$^-$, a constant factor applies to this ratio for different excitation powers. 

This, however, does not affect the relative trend in Fig. \ref{fig:plratio}, which shows the evolution of the $\mathrm{NV}^-/\mathrm{NV}^0$ ratio with increasing implantation fluence. Although the NV$^-$ and NV$^0$ emissions follow a similar non-monotonic trend with respect to the nitrogen saturation point at $\mathrm{F}=1\times10^{15}$ ions cm$^{-2}$, the $\mathrm{NV}^-/\mathrm{NV}^0$ ratio steadily decreases for increasing fluences, both in the low fluence and high fluence regimes. This behavior can be explained with the charge transfer process that determines the $\mathrm{NV}^0 \leftrightarrow \mathrm{NV}^-$ conversion. With increasing damage density, more nitrogen atoms pair with induced vacancies to form NV centers and fewer nitrogen donors are available to turn NV$^{0}$ centers into NV$^{-}$ centers. This imposes a natural limit to the $\mathrm{NV}^-$ density achievable with high energy He ion implantation for a given sort of diamond. As the presence of NV$^{0}$ is unfavorable for QIP, no more NV centers must be created than nitrogen atoms are available to provide the negative charge for it, i.e. in the case of full conversion efficiency, the vacancy density created should not exceed half  the density of nitrogen of the sample, leaving an electron donating nitrogen atom for each NV center formed. However, due to the limited mobility of the vacancies at the annealing temperature of our work, this saturation effect arises only slowly and at a higher vacancy densities of $3 \times 10^{20}$ vacancies cm$^{-3}$. 

\begin{figure}[tbp]	\includegraphics[width=0.95\columnwidth]{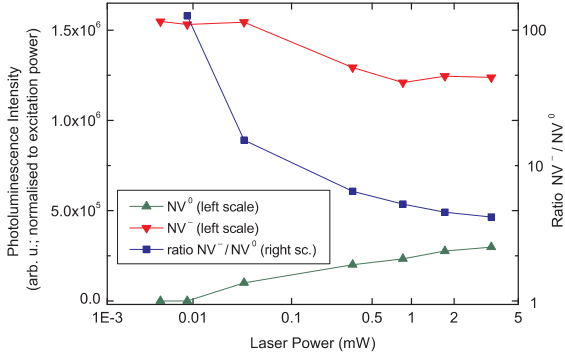}
	\caption{
	Excitation power dependence of the NV PL (normalized to excitation power) and PL intensity ratio (vacancy density $1.5 \times 10^{18}$ vacancies cm$^{-3}$). Photoionization leads to a strong increase in NV$^0$ signal for higher excitation powers. Ratio data points for which the NV$^0$ PL is not detectable have not been plotted.} 
	\label{fig:powerdepratio_r12}
\end{figure}

\subsection{Photochromism}
Since the attribution of the 575 nm emission to the neutral state of the NV defect by Mita \cite{PRB_53_11360}, conclusive work has demonstrated the photochromic behavior of the NV center  \cite{DRM_8_717,JPCM_12_189,axcm0508323}. Photoionization has been suggested as an explanation for photochromism, hypothesizing that an NV$^-$ center can be discharged by high power excitation light, resulting in an NV$^0$ center and a free electron. Fig. \ref{fig:powerdepratio_r12} shows the normalized PL intensity of NV$^0$ and NV$^-$ with respect to excitation power, for a constant ion implantation density ($\mathrm{F}=5 \times 10^{14}$ ions cm$^{-2}$). The normalized NV$^0$ PL intensity increases with increasing excitation power. At the powers used in our work ($\leq 5$ mW), PL saturation cannot be observed. For low excitation power, the NV$^0$ emission becomes negligible, as the vast majority of NV centers is negatively charged, whereas with increasing excitation power, the fraction of neutral NV centers increases. A decrease of the signal strength of NV$^-$ PL (normalized to laser power) occurs in the same way as NV$^0$ PL rises.  Assuming that the number of NV centers is proportional to the photoluminescence intensity multiplied by the radiative lifetime for each charge state (negative: 12 ns, neutral: 20 ns  \cite{zaitsev_book}), the total number of NV centers in the laser spot remains constant (within 4 \%) for the range of excitation powers probed. This derivation is only approximative, as it assumes the same excitation rate for both charge states, which is not necessarily the case at the given excitation wavelength. But it hints that conversion occurs only between the two charge states of the NV center, whereas optical conversion of the NV$^-$ center into non-radiative centers, if occuring at all, is minor.
With an excitation energy of 2.33 eV of our system, this ionization process can only occur via a resonant two-photon process by subsequent excitation and ionization of the NV$^-$ center (ground state 2.58 eV below conduction band \cite{scdg2000}). The build-up time for the NV$^0$ PL signal has been reported as on the order of hundreds of $\mu$s  \cite{DRM_14_1705}. At our peak excitation power, a conversion efficiency of up to 20 \% could be achieved, ranging similarly as reported by Manson \textit{et al.}  \cite{DRM_14_1705}. The photoionization conversion efficiency is limited by spontaneous back-reaction of a conduction band electron and NV$^0$ center into an NV$^-$ center, leading to an equilibrium for continuous-wave excitation. 

\begin{figure}[tbp]	
\includegraphics[width=1.0\columnwidth]{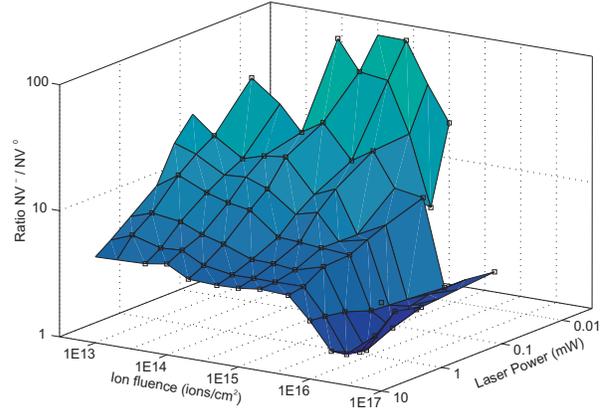}
	\caption{
	PL ratio of NV$^-$/NV$^0$ versus implantation fluence and laser excitation power (logarithmic scales). The figure shows clearly that a high ratio can be reached even for medium-high implantation densities, indicating temporal stability of the NV$^-$ centers in high-density ensembles. Data points for which the NV$^0$ PL is not detectable have not been plotted. In the figure, the surface has been obtained from the data points by spline interpolation. The black squares show the actual data points recorded. The PL linewidth (not plotted) remained constant at all laser powers within the accuracy of our detector (0.1 nm).
	}
	\label{fig:powerdepratio_all}
\end{figure}

As the charge conversion process implies a change of resonance frequency and a loss of the electron to the conduction band with subsequent fast intraband decoherence, QIP schemes involving single NV$^-$ emitters would most likely fail in the case of light induced charge conversion. Schemes that involve ensembles of NV centers suffer a fidelity decline from a partial charge conversion of the contributing centers. It is therefore vital for QIP with NV$^-$ ensembles to know the conditions for which the charge conversion of NV$^-$ can be minimized. For the off-resonant Qmem scheme mentioned above, it is worth noting that the collective coupling strength of a high density NV$^-$ ensemble allows for control laser powers much lower than the laser powers used here for excitation. Moreover, the two-photon charge transfer process is likely to be strongly suppressed for off-resonant frequencies. 

It is interesting to relate the photochromic behavior of the NV center to the ion implantation density. Fig. \ref{fig:powerdepratio_all} shows a systematic analysis of the NV$^-$/NV$^0$ ratio as a function of implantation fluence and excitation power. As discussed above, at high implantation densities, the fraction of negatively charged NV centers decreases due to a lack of free electrons. Therefore, for very high implantation densities the NV$^0$ emission cannot be suppressed and the temporal stability of the NV$^-$ center is not guaranteed. On the other hand, for lower implantation densities, the presence of NV$^0$ can be completely suppressed within the accuracy of our optical detection apparatus (at a signal to noise ratio for NV$^-$ PL of over 100). Our systematic analysis (figure \ref{fig:powerdepratio_all}) shows that such a regime can even be reached for medium ion implantation fluences of up to $5 \times 10^{14}$ ions cm$^{-2}$. Therefore, the technique of ion implantation of nitrogen rich HPHT diamond can be deployed to create high densities of NV$^-$ centers without a detectable impact on their temporal stability. 

\subsection{Inhomogeneous Broadening}
\begin{figure}[tbp]
	\includegraphics[width=1.0\columnwidth]{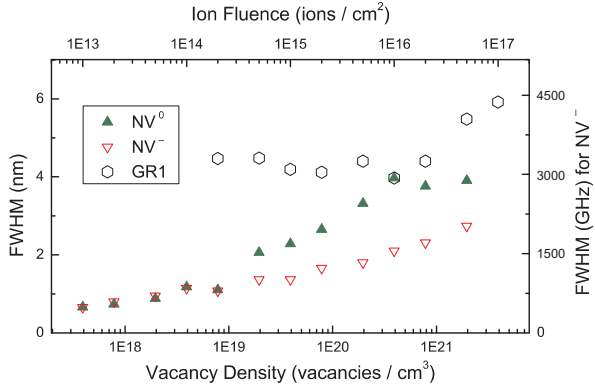}
	\caption{ PL peak widths of NV$^0$, NV$^-$ and GR1 with respect to irradiation fluence. An increase of the irradiation induced damage of the center environment leads to a stronger broadening of the luminescence line.} 
	\label{fig:regiondepwidth}
\end{figure}

The linewidth of the NV$^-$ PL peak in our sample is large, despite the cryogenic temperatures. This is typical for NV centers created by ion implantation in impurity-rich HPHT diamond. The peak widths measured (full width at half maximum, FWHM) monotonically increase from $\Delta \lambda = 0.66$ nm ($\Delta \nu \sim 480$ GHz) to $\Delta \lambda = 2.7$ nm ($\Delta \nu \sim 2$ THz) as a function of implantation fluence (see figure \ref{fig:regiondepwidth}). We attribute this increase to the implantation induced damage to the crystal structure and an increase in stress, which is not fully repaired due to the comparatively low annealing temperature. On the other hand, we could not detect a change of the NV$^-$ peak width due to a variation of the excitation power. This is the case for all implantation regions, such that illumination induced broadening does not occur at any extent of implantation induced damage. We therefore conclude that the increase of free charge carriers due to high power laser illumination ($1.6 \ {\mathrm{kW \ cm}}^{-2}$) has a small impact on the inhomogeneous linewidth compared to the broadening mechanisms prevalent in high nitrogen density diamond. 

A large inhomogeneous broadening of the NV$^-$ center ZPL has been reported for a great variety of samples \cite{zaitsev_book}. Recent research has shown a great interest in a reduction of this broadening, as a large inhomogeneous linewidth of an optical emitter limits its usability for certain QIP schemes. For example, a quantum memory that stores a photon in the excited state of an ensemble of NV$^-$ centers would dephase within picoseconds. Thus the controlled reversible inhomogeneous broadening (CRIB) scheme \cite{mk2001,nikr2005} could not be realized in practice without additional efforts, e.g. spectral hole burning. Likewise, an EIT Qmem would require huge control pulse energies to achieve a usable transparency window (fully separated absorption peaks), which implies intensities above the material damage threshold. It appears that, with current characteristics of high density NV samples, resonant Qmem schemes can only be applied to single NV centers, where inhomogeneous broadening is not present and spectral diffusion can be limited. For example, it has recently been shown \cite{sfsfbgodrrgrjp2006} that a great reduction of the nitrogen density by a careful choice of the raw sample material and deliberate implantation of single nitrogen molecules can lead to NV$^-$ linewidths of about $\Delta \nu \sim \unit[10]{GHz}$. While this technique can create single NV centers, it is unclear whether high density NV ensembles with as little inhomogeneous broadening can ever be produced. It has to be reminded that on-resonant schemes as the EIT scheme require equally high optical densities ($D \sim 10^3$) as the here presented off-resonant scheme  \cite{gals2007i,gals2007ii,gals2007iii}. Moreover, a linewidth of $\unit[10]{GHz}$ is still two orders of magnitude larger than the natural linewidth (83~MHz \cite{zaitsev_book}), such that for CRIB with spectral hole burning, the required optical density equally increases by this large factor. 

The off-resonant Qmem scheme therefore appears advantageous, as inhomogeneous broadening of the excited state $\ket{2}$ causes no difficulties. The major condition for off-resonant storage is a detuning much larger than the photon bandwidth and the excited state linewidth. Its dephasing time is subject to inhomogeneous broadening of the storage state only, not of the excited state. From the spectra recorded and using curve fitting, the inhomogeneous linewidth can be estimated as $\Delta \nu_{13} \sim \unit[2]{THz}$, which corresponds to a local phonon dephasing time of $\unit[500]{fs}$, which interestingly is a good order of magnitude shorter than the bulk phonon lifetime in diamond \cite{ihk2007,wlnsljwsop2008}. This would lead to rapid dephasing if this particular sample was used to build a PSB Qmem and the Qmem $Q$-factor would be low ($Q \equiv \Delta \nu_{13} \ \delta \sim 1.5$). However, no temporal analysis of the PSB lifetime has been published, such that the contributions of natural linewidth (leading to decoherence due to phonon decay) and inhomogeneous broadening (leading to dephasing) are not clear. The latter could possibly be minimised for improved sample creation methods. As mentioned above, the PSB storage scheme is generic and can be applied to other color centers, including those with longer lifetimes of the local phonon mode.

\section{Conclusion}
We have analyzed the photoluminescence properties of a type-Ib diamond sample, which has been systematically He implanted over a wide range of fluences. At low and medium fluences, the PL signal from NV centers increases with the fluence, whereas at high fluences absorption and competing processes lead to a decline of the NV luminescence. The ratio of NV$^-$/NV$^0$ fluorescence decreases for high implantation densities, as the density of residual nitrogen atoms which act as electron donors decreases. We have analyzed this ratio with respect to irradiation density and excitation power. The concurrent examination by ion implantation fluence and excitation power reveals suitable conditions for which the NV$^-$ center is a temporally stable absorber in the solid state, as the charge-conversion due to photoionization can be strongly suppressed. The results have been reviewed in the context of QIP. 

We have introduced a generic scheme to store broadband photons in ensembles of color centers in diamond via an off-resonant $\Lambda$-transition. A phonon sideband is used as the storage state, implying a short coherence time, which however is compensated by a large allowed bandwidth and ultrafast operation. Due to inhomogeneous broadening, the only practical quantum memory type suitable for NV \emph{ensembles} is the off-resonant (Raman) type. On-resonant schemes such as EIT are more practical to implement with \emph{single} NV centers, but require microcavities.

\section{Acknowledgments}
The authors gratefully acknowledge support by the {QIPIRC} and {EPSRC} (grant number {GR/S82176/01}), the Australian Research Council, the Australian government, the US National Security Agency (NSA), Advanced Research and Development Activity (ARDA), and the Army Research Office (ARO) under contract number W911NF-05-1-0284 and DARPA QUIST. FW thanks Toshiba Research Europe for their support. We thank Martin Castell, Andrew Briggs and Chris Salter supporting us in the analysis of the sample.

\bibliographystyle{unsrt}
\bibliography{nvpl}

\end{document}